\newcommand\ep{\epsilon}

\newcommand\fr{\frac}
\newcommand\pr{\prime}

\newcommand{\diracslash}[1]{#1\llap{/\kern2pt}}

\def\bearr{\begin{eqnarray}}

\def\eearr{\end{eqnarray}}

\newcommand{\be}{\begin{equation}}

\newcommand{\ee}{\end{equation}}

\newcommand{\bea}{\begin{eqnarray}}

\newcommand{\eea}{\end{eqnarray}}

\newcommand{\ba}[1]{\begin{array}{#1}}

\newcommand{\ea}{\end{array}}

\newcommand{\eqrf}[1]{Eq.\ (\ref{#1})}

\newcommand{\eqrftw}[2]{Eqs.\ (\ref{#1}) and (\ref{#2})}

\newcommand{\eqrftr}[3]{Eqs.\ (\ref{#1},\ref{#2},\ref{#3})}

\documentclass[preprint,secnumarabic,floats,amssymb,bibnotes,nofootinbib,showpacs,tightenlines]{revtex4}

\usepackage{graphicx}
\usepackage{fixltx2e}
\usepackage{amsmath}
\usepackage{multirow}
\usepackage{latexsym}

\usepackage{epsfig}

\usepackage{subfigure}

\begin{document}
\title{Reheating constraints on Tachyon Inflation}
\author{Akhilesh Nautiyal}
\affiliation{Department of Physics, Malaviya National Institute of Technology Jaipur, JLN Marg, Jaipur-302017}
\begin{abstract}
Tachyon inflation is one of the most attractive models of noncannonical inflation motivated by string theory.  In this work we 
revisit the constraints on tachyon inflation with inverse $\cosh$ potential and exponential potential considering reheating, 
Although the phase of reheating is not well understood,  it can be parameterized in terms of  reheating temperature $T_{re}$, 
number of e-folds during reheating $N_{re}$ and effective equation of state during reheating $w_{re}$, 
which can be related to the parameters of the tachyon potential, spectral index $n_s$ and tensor-to-scalar ratio $r$. 
For various reheating scenarios  there is a finite range  of $w_{re}$  and the reheating temperature should be above 
electroweak scale.  By imposing these conditions, we find that both the inverse $\cosh$ potential and exponential potential 
are disfavored by Planck observations. We also find that $w_{re}$ for both these potentials  should be close to $1$ to satisfy
Planck-2018 joint constraints on $n_s$ and $r$.  
\end{abstract}
\pacs{98.80.Cq, 14.80.Va, 98.80.-k,98.80.Qc}
\maketitle
\section{Introduction} Inflation \cite{Guth:1980zm} not only provides 
solution to the problems of big bang cosmology, but also generates seeds for 
CMB anisotropy and structures in the universe 
\cite{Mukhanov:1981xt,Starobinsky:1982ee,Guth:1985ya}. 
It predicts adiabatic, nearly 
scale invariant and Gaussian perturbations which are confirmed by various CMB  
observations like COBE \cite{Smoot:1992td}, WMAP \cite{Komatsu:2010fb}
and Planck \cite{Ade:2015lrj,Akrami:2018odb}. In standard scenario the potential energy of a 
scalar field, named as inflaton, dominates the energy density of universe for a
short period of time that causes the rapid expansion of the universe. The 
inflaton field $\phi$, whose dynamics can be determined by the Klein-Gordon 
action, rolls slowly through its potential. During this period the quantum
fluctuations in the scalar field, which are coupled to the metric fluctuations,
generate the primordial density perturbations. There are also vacuum 
fluctuations of the metric during inflation that generate primordial 
gravitational waves (tensor perturbations). 
The power spectra of the primordial density perturbations and tensor 
perturbations generated during inflation depend on the potential of the 
inflaton, which can be obtained from particle physics models or string theory. 

There are some alternative to the standard 
inflationary models \cite{ArmendarizPicon:1999rj,Garriga:1999vw}, named as 
K-inflation, where the inflation is achieved by non-standard kinetic term in 
the Lagrangian of the inflaton. One attractive and popular model of 
K-inflation is tachyon inflation \cite{Gibbons:2002md}, which can be realized 
in Type-II string theory where the tachyon signals the instability of unstable
and uncharged D-branes of tension $\lambda$. In this case the tachyon action
is of the Dirac-Born-Infield form (see \cite{Sen:1999md,Garousi:2000tr,
Bergshoeff:2000dq,Kluson:2000iy,Sen:2002qa,Kutasov:2003er,Okuyama:2003wm} for different approaches).  
It was also pointed out by Sen \cite{Sen:2002nu} that the rolling tachyon 
can have the dust like equation of state, which raised the possibility of 
tachyon providing a unified description of inflation and dark matter. Tachyon as inflaton has been 
criticized in \cite{Frolov:2002rr,Kofman:2002rh} as the string theory motivated values of the parameters are 
incompatible with the slow-roll conditions and observed amplitude of the 
scalar perturbations. However, there are some scenarios \cite{Piao:2002vf,Mazumdar:2001mm,Chingangbam:2004ng,Choudhury:2015hvr} 
where  tachyon as inflaton can satisfy the 
observations with parameters obtained from string theory. Despite the criticism and regardless the string theory 
motivation, tachyon inflation has been studied phenomenologically as an example
of noncannonical inflation model \cite{Steer:2003yu,Barbosa-Cendejas:2017pbo}.
In \cite{Steer:2003yu} it was shown that the consistency relation in case of 
tachyon inflation is different than the standard single field inflation and an
observational signature of this deviation can lead to distinguish between the 
two models. 

Inflation leaves the universe in a cold and highly nonthermal state without any matter content. 
Universe needs to be in a thermalized state at a very high temperature for hot big bang picture. 
This is achieved by reheating which is   a transition phase between the end of inflation and the start of radiation 
dominated era. During reheating the energy density of inflaton is 
converted  to the thermal bath, at a reheating temperature $T_{re}$, that fills
the universe at the beginning of radiation dominated era 
(see \cite{Allahverdi:2010xz} for detailed review). In simplest case 
reheating can occur via perturbative decay of inflaton into standard model 
matter particles, while inflaton is oscillating around the minimum of its 
potential \cite{Abbott:1982hn,Dolgov:1982th,Albrecht:1982mp}, but  this 
scenario was criticized in  \cite{Traschen:1990sw,Dolgov:1989us} as it 
does not take into account the coherent nature of the inflaton field.  In other scenarios reheating is
preceded by phase of preheating during which the particle production occurs 
via non-perturbative  processes such as 
parametric resonance decay \cite{Kofman:1994rk,Kofman:1997yn}, 
tachyonic instability \cite{Greene:1997ge, Dufaux:2006ee} and instant 
preheating \cite{Felder:1998vq}. Preheating leaves the universe in highly
nonthermal state which is thermalized by scattering and the universe is left
with a blackbody spectrum at a temperature $T_{re}$, named as reheating 
temperature, which corresponds to  the temperature at the beginning of the 
radiation dominated epoch. It is difficult to constrain the reheating 
temperature $T_{re}$ from CMB and LSS observations, but it is considered that $T_{re}$ should be above the electroweak scale: 
so that the weak scale dark matter can be produced. If we adopt a conservative approach, $T_{re}$ must be greater than $10$MeV
for the big bang nucleosynthesis.  
If late-time entropy production by massive particle decay is considered \cite{Kawasaki:1999na,Kawasaki:2000en}, 
the reheating temperature can be as 
low as $2.5$ to $4$MeV.
The upper bound on reheating temperature is obtained by assuming reheating to be an instantaneous
process which reheats the universe to the scale of inflation i.e. $10^{16}$GeV by considering the Planck upper bound on 
tensor-to-scalar ratio. The evolution of the energy density of the cosmic fluid 
during reheating depends on effective equation of state $w_{re}$, which, in general, depends on time. 
Its value at the end of inflation is $-\frac13$ and reaches $\frac13$ at the beginning of radiation dominated epoch. In case where
the reheating occurs due to perturbative decay of massive inflaton, the effective equation of state during reheating $w_{re}$ 
is $w_{re}=0$ and for instant reheating $w_{re}=\frac13$. 
A numerical study performed for various reheating scenario \cite{Podolsky:2005bw} shows that $w_{re}$ can vary between $0$ to 
$0.25$. Another important parameter to describe reheating is its duration, which can be defined in terms of number of e-foldings
$N_{re}$ from the end of inflation to the onset of radiation dominated epoch. In general, this is incorporated by providing 
a range for $N_k$, the number of e-folds from the time when a Fourier mode $k$ corresponding the the horizon size of our
 observable universe  leaves the inflationary horizon to the end of 
inflation. $N_k$ depends on the potential of inflaton and it should be between $46$ to $70$ to solve the horizon problem. The 
upper bound on $N_k$ comes from assuming instantaneous reheating and the lower bound arises from considering reheating 
temperature at electroweak scale. A detailed analysis of upper bound on $N_k$ for various scenarios was presented in 
\cite{Liddle:2003as,Dodelson:2003vq} and it was shown that for some cases $N_k$ can be as large as $107$.

These three parameters of reheating can be used to obtain constraints on  various inflationary models 
\cite{Dai:2014jja,Munoz:2014eqa,Cook:2015vqa}. 
Demanding that the equation of state during reheating lies between $0$ and $0.25$, one can get  bounds on the 
spectral index $n_s$ and $N_k$, which translates to bounds on tensor-to-scalar ratio $r$. In this work we use this approach to constrain
tachyon inflation with inverse $\cosh$ and exponential potentials.  
We obtain $T_{re}$ and $N_{re}$ as a function of spectral index as in \cite{Cook:2015vqa} by assuming $w_{re}$ to be 
constant during reheating. We obtain the allowed regions for these potentials in $n_s-r$ plane for various values of $w_{re}$. 
We also use Planck-2018 $1\sigma$ bounds on $n_s$ and $r$ \cite{Akrami:2018odb} 
and BICEP2/Keck Array bounds on $r$ \cite{Ade:2018gkx}  
to determine the equation of state during reheating for these 
potentials. 

The work is organized as follows: in Sec.~\ref{TI} we give a brief review of tachyon inflation providing the 
expressions for $N_k$, scalar power spectrum, spectral index $n_s$ and tensor t0 scalar ratio $r$. In 
Sec.~\ref{reheating} we provide the derivation for reheating temperature $T_{re}$ and number of e-folds during reheating 
$N_{re}$ for constant effective equation of state $w_{re}$. In Sec.~\ref{cti} we obtain  $T_{re}$, and $N_{re}$ for tachyon 
inflation with inverse $\cosh$ and exponential potential for various choices of equation of state $w_{re}$ and use these three
parameters to constrain tachyon inflation. In Sec.~\ref{conclusions} we conclude our work.

\section{Tachyon Inflation}\label{TI}
Tachyon inflation is a class of $K$-inflation models where inflation is achieved by noncannonical kinetic term. The action for 
tachyon inflation is given by
\be
S_T=-\int d^4x\sqrt{-g}V(T)\left(1+g^{\mu\nu}\partial_\mu T \partial_\nu T\right)^{\frac12}, \label{tachyaction}
\ee
and the metric has signature $-$,$+$,$+$,$+$. $T$ represents the tachyon field with dimension of length and $V(T)$ represents
its potential. Various choice for the potential have been derived using string theory \cite{Sen:1999md,Garousi:2000tr,Bergshoeff:2000dq,Kluson:2000iy,Sen:2002qa,Kutasov:2003er,Okuyama:2003wm}. Here we consider the inverse $\cosh$ potential 
\cite{Sen:1999md,Garousi:2000tr,Lambert:2003Ar} 
 and exponential potential \cite{Sen:2002an,Sami:2002fs} given by
\be
V(T)=\frac{\lambda}{\cosh\left(\frac{T}{T_0}\right)}, \label{coshpotential}
\ee
and 
\be
V(T)=\lambda\exp\left(-\frac{T}{T_0}\right). \label{exppotential}
\ee

The action for the tachyon-gravity system is given by 
\be
S= \int d^4x \sqrt{-g}\frac{R}{16\pi G}+S_T. \label{fullaction}
\ee
The energy-momentum tensor can be obtained by varying this action as 
\be
T_{\mu\nu}=-V(T)g_{\mu\nu}\sqrt{1+\partial^\mu T\partial_\mu T}+\frac{V(T)}{\sqrt{1+\partial^\mu T\partial_\mu T}}
\partial_\mu T \partial_\nu T. \label{tmunu}
\ee
The energy density and pressure for the background part of the tachyon field in a homogeneous and isotropic universe is given 
as
\bea
\rho &=& \frac{V(T)}{\left(1-\dot T^{2}\right)^{\frac12}}, \label{energydensity} \\
p&=&-V(T)\left(1-\dot T^{2}\right)^{\frac12}.\label{presure}
\eea
Hence the Friedmann equation becomes
\be
H^2=\frac{1}{3M_p^2}\frac{V(T)}{\left(1-\dot T^{2}\right)^{\frac12}}, \label{hubbletachy}
\ee
and the equation of motion for the background part of the tachyon field can be obtained using the conservation of 
energy-momentum tensor as
\be
\frac{\ddot T}{\left(1-\dot T^2\right)}+3 H \dot T+\left(\ln V\right)^\prime =0. \label{eqomtachy}
\ee 
The conditions to achieve inflation can be obtained by using Friedmann equation
\be
\frac{\ddot a}{a} = -\frac{1}{6 M_p^2}\left(\rho+3p\right)=\frac{1}{3M_p^2}\frac{V}{\left(1-\dot T^2\right)^{\frac12}}
\left(1-\frac{3}{2}\dot T^2\right)>0, \label{conditioninflation}
\ee
which gives $\dot T^2<\frac32$. And also for inflation to last sufficiently longer $\ddot T$ should be smaller than the friction
term in the equation of motion for tachyon field \eqrf{eqomtachy}
\be
\ddot T<3 H \dot T. \label{conditiontwo}
\ee
Hence during inflation
\be
\dot T \sim - \fr{\left(\ln V\right)^\pr}{3H}, \, \, H^2 \sim \frac{V}{3M_p^2}. \label{condition3}
\ee
To analyze the dynamics of inflation the slow-roll parameters can be defined in terms of the Hubble flow parameters
\cite{Schwarz:2001vv} as 
\bea
\epsilon_0 &\equiv& \frac{H_k}{H} \label{epzero}, \\
\epsilon_{i+1}&\equiv&\fr{d\ln|\ep_i|}{dN}, \, \, \, i\ge 0, \label{epi}
\eea
where $H_k$ is the Hubble constant during inflation when a particular mode $k$ leaves the horizon and $N$ is the number of 
e-foldings
\be
N\equiv \ln\left(\frac{a}{a_i}\right) \label{efolds1},
\ee
where $a_i$ is the scale factor at the beginning  of inflation. We also have
\be
\dot \ep_i = \ep_i\ep_{i+1}. 
\ee
In terms of $T$ the slow-roll parameters defined by \eqrftw{epzero}{epi} can be written as
\bea
\ep_1&=&\fr32 \dot T^2, \label{ep1}\\
\ep_2&=&\sqrt{\fr{2}{3\ep_1}}\fr{\ep^\pr}{H}=2\frac{\ddot T}{H\dot T}.\label{ep2}
\eea
For conditions (\ref{condition3}) to be satisfied, $\ep_1,\ep_2\ll 1$ and inflation ends when $\ep_1=1$.

The power spectra for scalar and tensor perturbations, spectral index and tensor to scalar ratio is these models are given as
\cite{Garriga:1999vw,Steer:2003yu}
\bea
P_\zeta&=&\left.\frac{H^2}{8\pi^2M_p^2c_S\epsilon}\right|_{c_S k=aH},\label{scalaramp}\\
P_h&=&\left.\frac{2}{\pi^2}\frac{H^2}{M_p^2}\right|_{k = aH }, \label{tensoramp}\\
n_s&=&1-2\ep_1-\ep_2, \label{ns}\\
r&=&16c_S \ep. \label{tsratio}
\eea
Here $c_S$ is the effective sound speed given as 
\be
c_S^2=\frac{\partial P/\partial \dot T^2}{\partial \rho/\partial \dot T^2} = 1-\dot T^2. \label{soundspeed}
\ee
The effective sound speed for these models is very close to $1$. The power spectrum $P_\zeta$, spectral index $n_s$ and 
tensor-to-scalar ratio $r$ are all evaluated at the pivot scale $k=k_0$ which is $0.05$Mpc\textsuperscript{-1} for Planck observations.
All these parameters depend on the choice of the tachyon potential $V(T)$ and are constrained by CMB and LSS observations. In 
Sec.~\ref{cti}, we will discuss how reheating can be used to limit our choice of the potentials. Before that we discuss the 
relation between reheating parameters and inflationary parameters in the next section.
\section{Reheating}\label{reheating}
Models of reheating can be parameterized in terms of thermalization temperature $T_{re}$ at the end of reheating, 
duration of reheating $N_{re}$ and equation of state during reheating $w_{re}$ \cite{Munoz:2014eqa,Cook:2015vqa}. 
We consider $w_{re}$ to be constant during reheating. The value of $w_{re}$ should be larger 
than $-\frac{1}{3}$ for inflation to end and it should be smaller than $1$ to satisfy dominant energy condition 
of general relativity, $\rho \ge |p|$, for the causality to be preserved \cite{Munoz:2014eqa,Martin:2010kz,Chavanis:2014lra}.
   
If the equation of state remains the same during reheating, the change in the scale factor can be related to the change
in energy density by using $\rho= a^{-3(1+w)}$ as
\be
\frac{\rho_{end}}{\rho_{re}}=\left(\frac{a_{end}}{a_{re}}\right)^{-3\left(1+w_{re}\right)}. \label{are}
\ee
Here the subscripts {\it end} and {\it re} denote the values of the quantity at the end of inflation and at the end of
reheating respectively. \eqrf{are} can be expressed in terms of $N_{re}=\ln\left(\frac{a_{re}}{a_{end}}\right)$ as
\bea
N_{re} &=& \frac{1}{3\left(1+w_{re}\right)}\ln\frac{\rho_{end}}{\rho_{re}}\nonumber\\
&=&\frac{1}{3\left(1+w_{re}\right)}\ln\left(\frac32\frac{V_{end}}{\rho_{re}}\right).\label{nre1}
\eea 
 Here we have substituted $\rho_{end}=\frac32 V_{end}$ as the equation of state at the inflation is $-\frac13$. 
The relation between the reheating temperature $T_{re}$ and $N_{re}$ can be obtained by expressing the energy density at the 
end of reheating $\rho_{re}$ in terms of $T_{re}$ as  
\be
\rho_{re}=\frac{\pi^2}{30}g_{re}T_{re}^4. \label{rhoretre}
\ee
So from \eqrf{nre1} we get 
\be
N_{re}=\frac{1}{3\left(1+w_{re}\right)}\ln\left(\frac{30\cdot \frac{3}{2}V_{end}}{\pi^2g_{re}T_{re}^4}\right).\label{nre2}
\ee
 Using entropy conservation the temperature at the end of reheating can be related to the CMB temperature today as
\be
T_{re}=T_{0}\frac{a_0}{a_{re}}\left(\frac{43}{11 g_{re}}\right)^{\frac13}=T_0\frac{a_0}{a_{eq}}e^{N_{RD}}
\left(\frac{43}{11 g_{re}}\right)^{\frac13},\label{tret0}
\ee 
where $T_0$ is the CMB temperature today,  $a_0$ is the scale factor today,  $N_{RD}$ is the number of e-foldings during radiation
dominated epoch and $a_{eq}$ is the scale factor at matter-radiation equality. The ratio $\frac{a_0}{a_{eq}}$ can be expressed as
\be
\frac{a_0}{a_{eq}}= \frac{a_0}{a_k}\frac{a_k}{a_{end}}\frac{a_{end}}{a_{re}}\frac{a_{re}}{a_{eq}}=\frac{a_0 H_k}{k}
e^{-N_k}e^{-N_{re}}e^{-N_{RD}},\label{a0aeq}
\ee
where $a_k$ and $H_k$ are the values of scale factor and the Hubble constant during inflation when the Fourier mode $k$ leaves
the horizon,  $N_k$ is the number of e-foldings from this time to the end of inflation and $k=a_k H_k$ for horizon exit. 
Now using \eqrftw{a0aeq}{tret0} the relation between $T_{re}$ and $N_{re}$ can be expressed as
\be
T_{re}=T_{0}\frac{a_0H_k}{k}\left(\frac{43}{11g_{re}}\right)^{\frac13}e^{-N_k}e^{-N_{re}}.\label{trenre}
\ee
Substituting this into \eqrf{nre2} we obtain the expression for $N_{re}$ as
\be
N_{re}=\frac{4}{3\left(1+w_{re}\right)}\left(\frac14\ln\left(\frac{3^2\cdot 5}{\pi^2g_{re}}\right)+
\ln\left(\frac{V_{end}^{\frac14}}{H_k}\right)+\ln\left(\frac{k}{T_0a_0}\right)+
\frac13\ln\left(\frac{11g_{re}}{43}\right)+N_k+N_{re}\right).\label{nre3}
\ee
If we consider $w_{re}\ne\frac13$, \eqrf{nre3} can be solved to obtain $N_{re}$ as
\be
N_{re}=\frac{4}{1-3 w_{re}}\left(-\frac14\ln\left(\frac{3^2\cdot 5}{\pi^2g_{re}}\right)-\frac13\ln\left(\frac{11g_{re}}{43}\right)
-\ln\left(\frac{V_{end}^{\frac14}}{H_k}\right)-\ln\left(\frac{k}{T_0a_0}\right)-N_k\right).\label{nrefinal}
\ee 
For $w_{re}=\frac13$ reheating occurs instantaneously leaving the universe at grand unification scale and hence parameters of 
reheating cannot be used to constrain models of inflation. Using \eqrftw{nrefinal}{trenre}  the temperature at the end of reheating $T_{re}$ can be expressed as
\be
T_{re} = \left(\left(\frac{43}{11g_{re}}\right)^{\frac13}\frac{a_0T_0}{k}H_ke^{-N_k}
\left(\frac{3^2\cdot 5V_{end}}{\pi^2g_{re}}\right)^{-\frac{1}{3\left(1+w_{re}\right)}}\right)^
{\frac{3\left(1+w_{re}\right)}{3w_{re}-1}}.\label{trefinal}
\ee
The main results of this section are expressions for reheating temperature $T_{re}$ \eqrf{trefinal} and number of e-folds 
during reheating \eqrf{nrefinal} that depend on the inflationary parameters $H_k$, $N_k$ and $V_{end}$. In next section
we obtain these parameters for tachyon inflation with inverse $\cosh$ and exponential potential in terms of amplitude of scalar
perturbations $A_s$ and spectral index $n_s$. With this  $T_{re}$ and $N_{re}$ can be expressed as function of $n_s$ and are used
  to constrain tachyon inflation by demanding $w_{re}$ between $-\frac{1}{3}$ and $1$.
\section{Constraints on tachyon inflation} \label{cti}
 In this section we constrain tachyon inflation with inverse $\cosh$ (\ref{coshpotential}) and exponential (\ref{exppotential})
potential from reheating.  To simplify our calculations we define 
$x\equiv\frac{T}{T_0}$ and a constant dimensionless ratio 
$X_0^2\equiv\frac{\lambda T_0^2}{M_{pl}^2}$.  
\subsection{Inverse $\cosh$ potential}
The inverse $\cosh$ potential (\ref{coshpotential}) can be  obtained fro string theory 
\cite{Sen:1999md,Garousi:2000tr,Lambert:2003Ar} and is the most popular choice for tachyon potential. In terms of $x$ it 
can be written as
\be
V= \frac{\lambda}{\cosh x} \label{coshpt}
\ee
The two slow-roll parameters $\epsilon_1$ and $\epsilon_2$ for this model can be obtained using \eqrf{condition3}, \eqrf{ep1}
and \eqrf{ep2} as
\bea
\epsilon_1&=&\frac{1}{2X_0^2}\frac{\sinh^2x}{\cosh x}\label{ep1cosh}\\
\epsilon_2&=&\frac{1}{X_0^2}\frac{\cosh^2x+1}{\cosh x}\label{ep2cosh}
\eea 
We can obtain the value of the tachyon field at the end of inflation by putting $\epsilon_1=1$ and it gives
\be
\cosh x_{end}=X_0^2+\sqrt{X_0^4+1}\label{xend},
\ee
which gives $x_{end}=\ln 4 X_0^2$ for $X_0>1$. The number of e-foldings $N_k$ during inflation  from the time when mode
$k$ leaves the horizon to the end of inflation can be obtained using \eqrf{condition3} as
\bea
N_k&=&\int_{t_k}^{t_{end}}H(t)dt=\int_{T_k}^{T_{end}}\frac{H}{\dot T}dT\nonumber\\
&=&-\frac{1}{M_p^2}\int_{T_k}^{T_{end}}\frac{V^2}{V^\prime}dT, \label{efoldsT}
\eea 
which for the inverse $\cosh$ potential becomes
\be
N_k=X_0^2\int_{x_k}^{x_{end}}\frac{1}{\sinh x}dx
=X_0^2\ln\left(\frac{\tanh\frac{x_{end}}{2}}{\tanh\frac{x_k}{2}}\right)\label{efoldscosh}
\ee
Using \eqrf{xend} and $X_0>1$ it can be shown that $\tanh x_{end} \sim 1$ so the value of tachyon field at the time when
mode $k$ leaves the inflationary horizon can be given as
\be
\tanh \frac{x_k}{2}=e^{-\frac{N_k}{X_0^2}}. \label{xknk}
\ee
From \eqrf{xknk} we obtain
\be
\sinh x_k = \frac{1}{\sinh \left(\frac{N_k}{X_0^2}\right)},\, \, \cosh x_k = \frac{1}{\tanh\left(\frac{N_k}{X_0^2}\right)}.
 \label{xknk1}
\ee

The scalar power spectrum $P_\zeta$ (\ref{scalaramp}), spectral index $n_s$ (\ref{ns}) and tensor-to-scalar ratio 
(\ref{tsratio}) 
are evaluated at the horizon crossing $c_S k=a H$ for pivot scale $k=k_o$, which we choose  $0.05$Mpc\textsuperscript{-1} 
as in Planck. $P_\zeta$ at $k_0$ is equal to the amplitude of scalar perturbations $A_s$ so we can express the 
 Hubble constant $H_k$ using \eqrf{scalaramp} as
\be
H_k= \pi M_p \sqrt{8 A_s \epsilon_1 c_S}.
\ee
The spectral index (\ref{ns}) for the potential \eqrf{coshpt} can be obtained using \eqrftw{ep1cosh}{ep2cosh} as
\be
n_s=1-\frac{2}{X_0^2}\cosh x_k.\label{nscosh}
\ee
The  tachyon potential at the end of inflation can be obtained as
\be
V_{end}=\frac{\lambda}{\cosh x_{end}}=3M_p^2 H_k^2\frac{\cosh x_k}{\cosh x_{end}}
\ee
Using \eqrftr{condition3}{xend}{nscosh} we get
\be
V_{end} = \frac{3}{4}M_p^2 H_k^2\left(1-n_s\right).\label{vendfinal}
\ee
We can also write $N_k$ in terms of $n_s$ using \eqrftw{xknk1}{nscosh} as
\be
N_k=X_0^2 \tanh^{-1}\left(\frac{2}{X_0^2\left(1-n_s\right)}\right).\label{nkns}
\ee
Both the slow-roll parameters and $c_S$ can be expressed in terms of spectral index $n_s$ and so the Hubble constant $H_k$ can 
also be expressed in terms of $n_s$ as
\be
H_k=2\pi M_p\frac{1}{X_0} \sqrt{A_s\left(\frac{\frac12\left(1-n_s\right)}{X_0^2}-\frac{2}{\left(1-n_s \right)X_0^2}
\right)\left(1-\frac{\frac12\left(1-n_s\right)X_0^2-\frac{2}{\left(1-n_s\right)X_0^2}}{6X_0^2}\right)}\label{hkfinal}
\ee
It can be seen from \eqrftr{vendfinal}{hkfinal}{nkns} that $V_{end}$, $H_k$ and $N_k$ are all expressed in terms of 
amplitude of scalar perturbations $A_s$ and spectral index $n_s$. Hence  $T_{re}$ and $N_{re}$ can be obtained as a function
of $A_s$ and $n_s$  by putting these expressions in \eqrftw{nrefinal}{trefinal}. 
 We use Planck-2018 values  \cite{Akrami:2018odb} for $A_s=2.20\times 10^{-9}$ (central value) and 
$n_s=0.9653 \pm 0.0041$ for our analysis. The small error bars on $A_s$ does not affect the results.

\begin{figure}[h!]
 \centering
\subfigure{\includegraphics[width=7cm, height = 7cm]{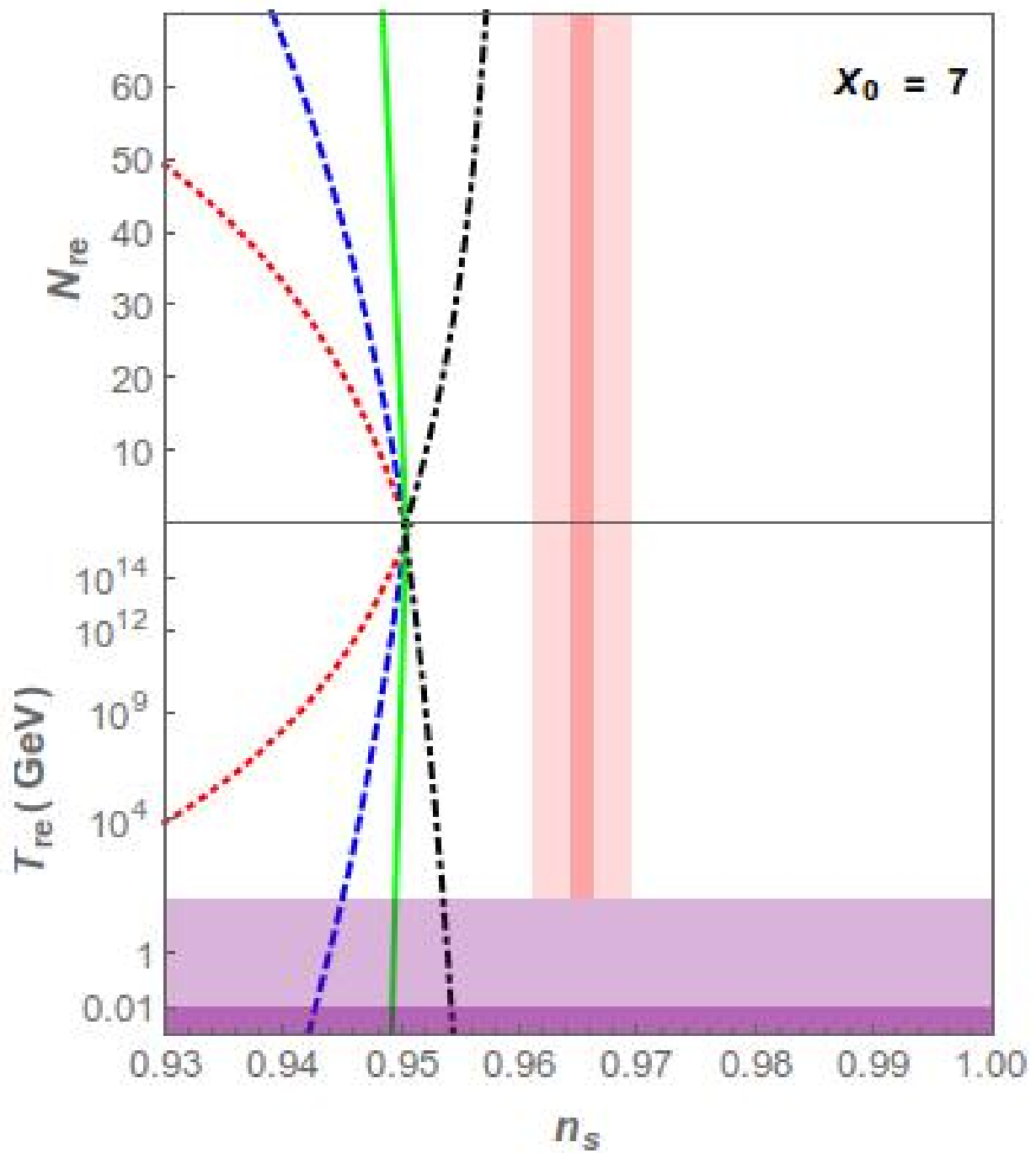}}
\subfigure{\includegraphics[width=7cm, height = 7cm]{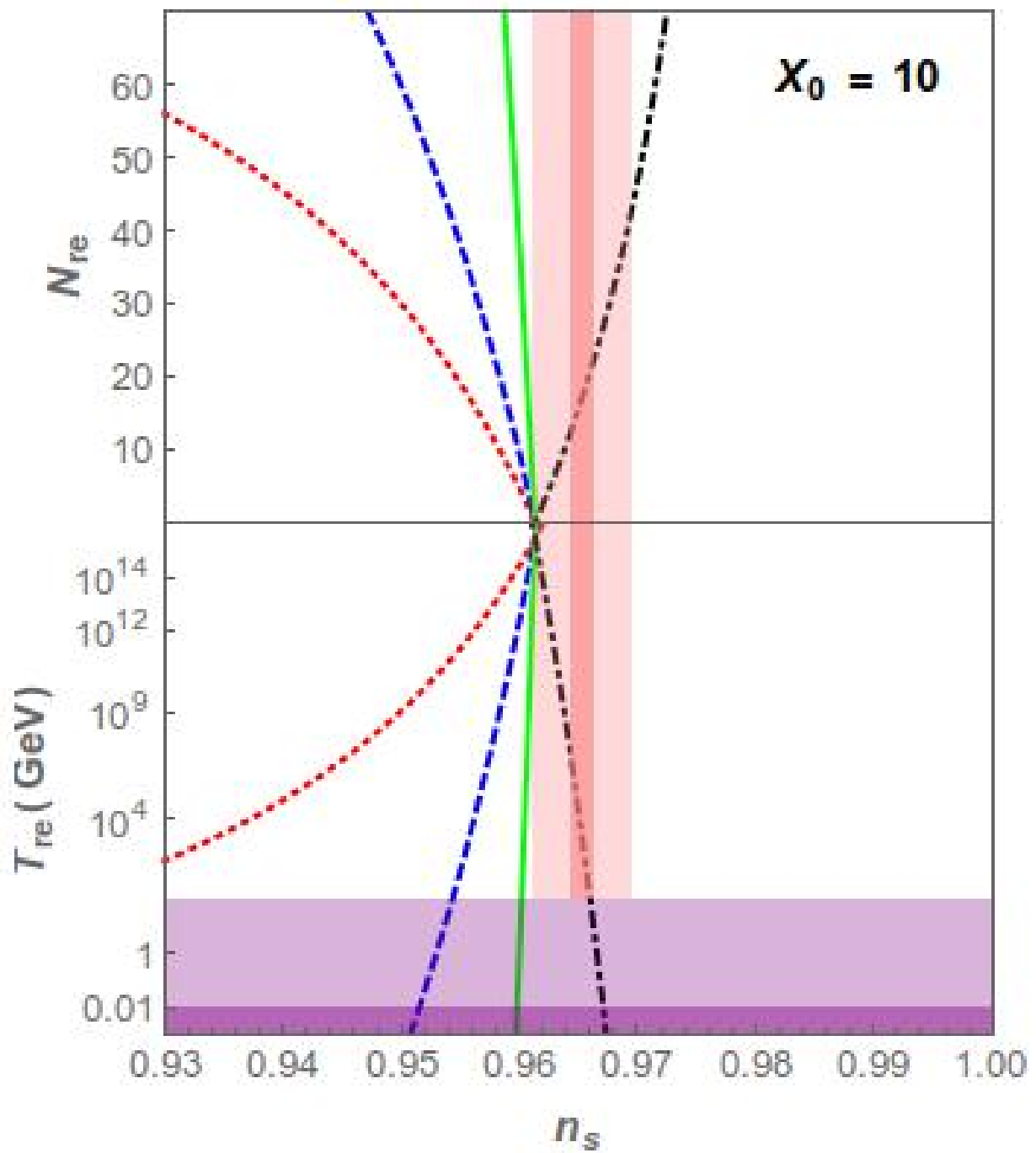}}
\subfigure{\includegraphics[width=7cm, height = 7cm]{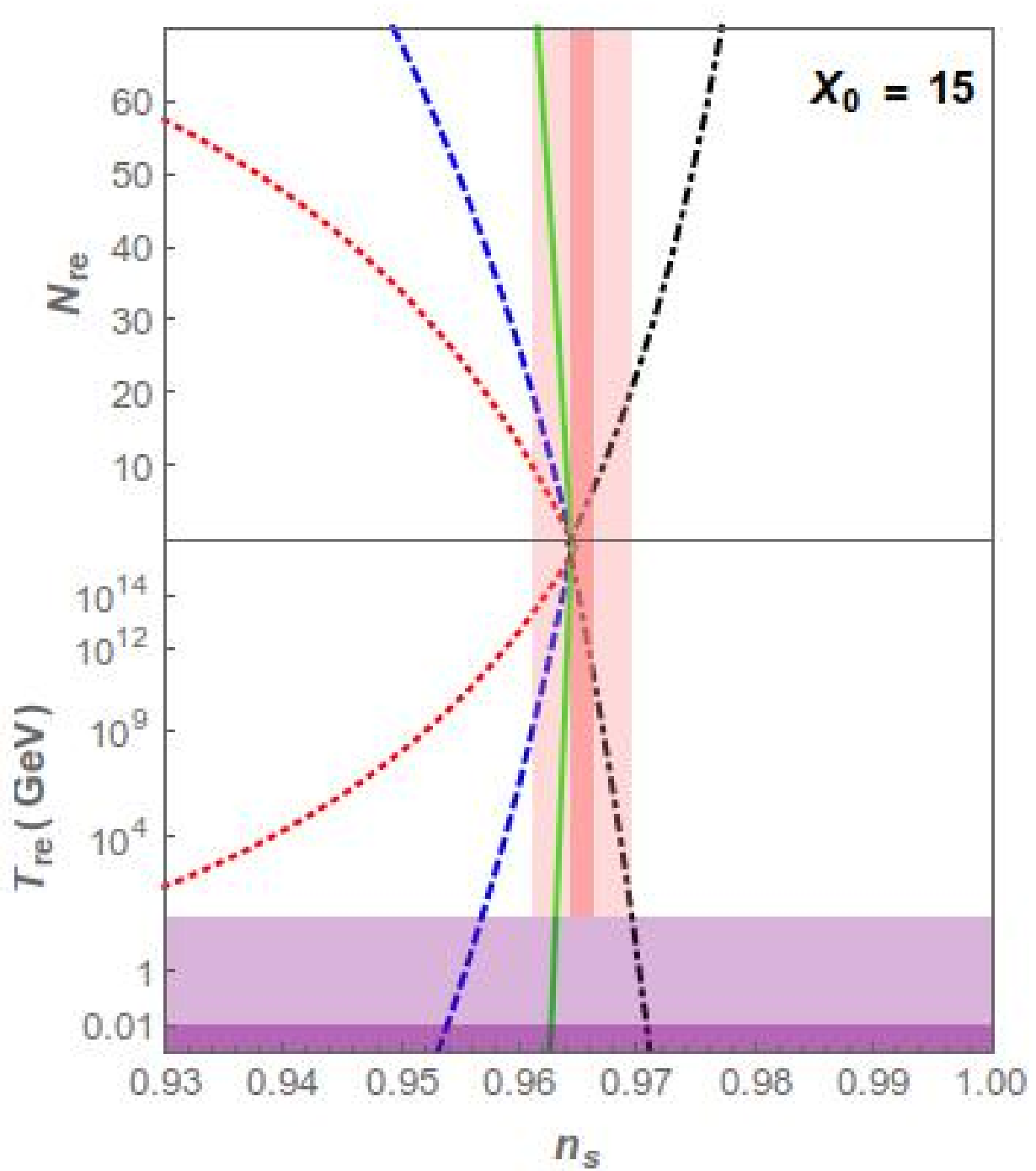}}
\subfigure{\includegraphics[width=7cm, height = 7cm]{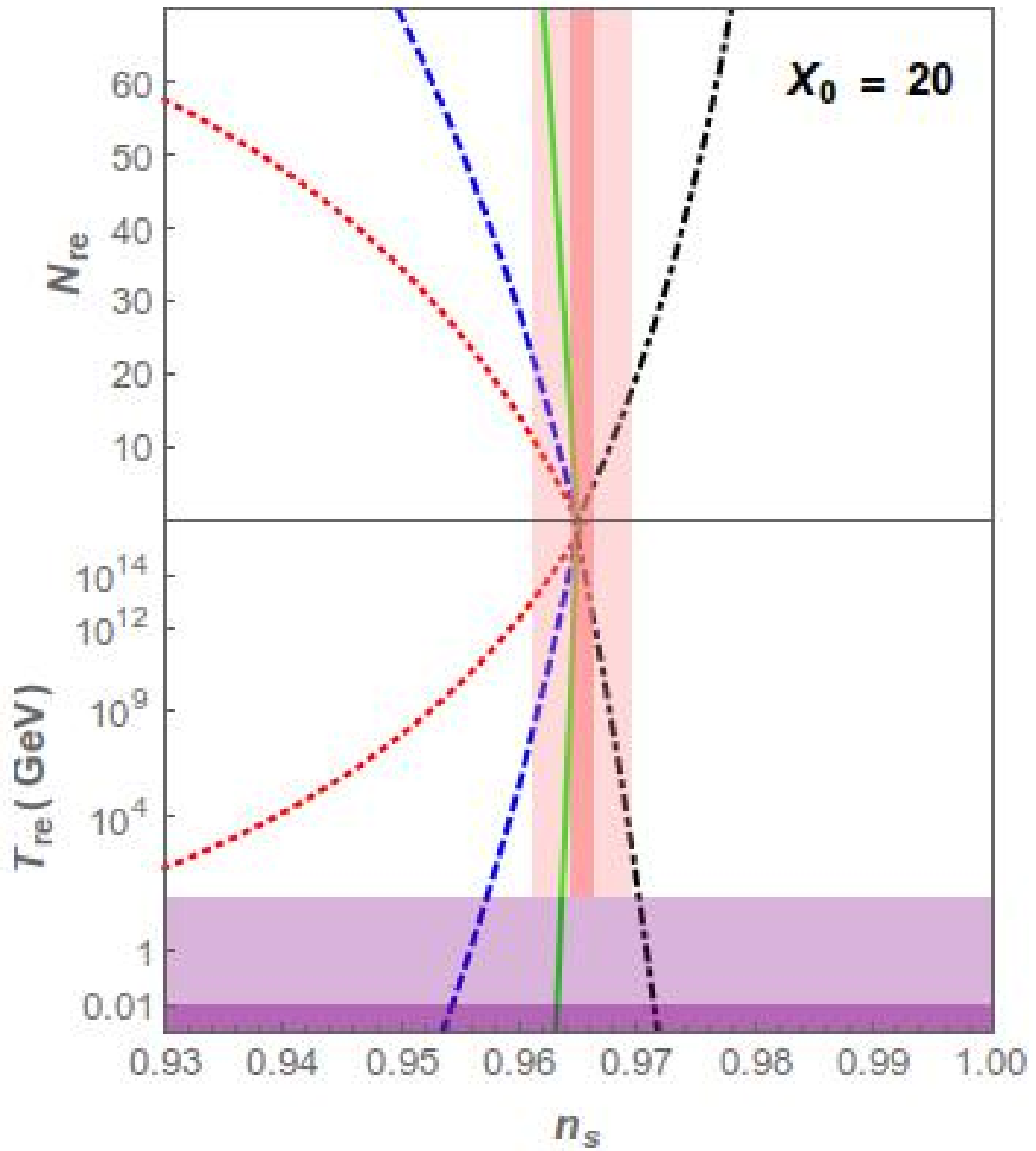}}
\caption{Figure shows $N_{re}$ and $T_{re}$, the length of reheating and temperature at the end of reheating respectively, 
as a function of $n_s$ for three different values of $X_o$ for inverse $\cosh$ 
potential. Here vertical light pink region represents Planck-2018 bounds on 
$n_s$ \cite{Akrami:2018odb} and dark pink region represents a precision of $10^{-3}$ from future experiments \cite{Amendola:2016saw}. 
Horizontal dark purple region 
represents $T_{re}$  of $10$MeV from BBN and light purple region represents $100$GeV of electroweak scale. Red dotted line 
corresponds to $w_{re}=-\frac{1}{3}$, blue dashed line corresponds to $w_{re}=0$, green solid line corresponds 
to $w_{re} = 0.25$ and black 
dotdashed line corresponds to $w_{re}=1$. For $X_0=7$ both $N_{re}$ and $T_{re}$ lie outside the Planck-2018 bound. }
\label{fig:coshnretre}
\end{figure}

 Fig.~\ref{fig:coshnretre} shows the variation of temperature at the end of reheating $T_{re}$ and the number of e-folds
during reheating $N_{re}$ as a function of spectral index $n_s$. We chose four values of $w_{re}$ between $-\frac13$ to $1$.
The curves for all $w_{re}$ meet at a point that corresponds to $w_{re}=\frac13$, which is defined as instant reheating 
($N_{re}\rightarrow 0$). The curve of $w_{re}=\frac13$ would pass through this point and be vertical.  
As depicted in the figure for $X_0=7$ the values of $T_{re}$ and $N_{re}$, for all choices of $w_{re}$, completely lie outside
the Planck-2018 bounds on $n_s$. So to satisfy the observations $X_0>1$, which justifies our assumption used in our calculations. 
We chose physically plausible values for $w_{re}$ .i.e. $0\le w_{re} \le 0.25$ obtained in \cite{Podolsky:2005bw} and demand that 
the reheating temperature $T_{re}$ should be larger than  $100$GeV (shown by light purple region in 
Fig.~\ref{fig:coshnretre}) for production of weak scale dark matter. This gives bounds on $n_s$ which are stronger than the 
Planck $1\sigma$ bounds (shown by light pink region in Fig.~\ref{fig:coshnretre}) for large value of $X_0$. These bounds on $n_s$ 
 correspond to the bounds on $N_k$, which can be obtained using \eqrf{nkns} and  are listed in Table \ref{table:coshnsnk}.

\begin{table} 
\begin{tabular}{|c|c|c|c|}
\hline
$X_0$ & Equation of state during reheating & $n_s$ & $N_k$\\
\hline
\multirow{2}{*}{$7 $} & $0 \le w_{re}\le 0.25$ & $0.945\le n_s\le 0.949$ & $46.4\le N_k\le 54.7$\\
\cline{2-4}
 & $0.25 \le w_{re}\le 1$ & $0.949\le n_s\le 0.954$ & $54.7\le N_k\le 67.0$\\
\hline
\multirow{2}{*}{$10 $} & $0 \le w_{re}\le 0.25$ & $0.954\le n_s\le 0.959$ & $46.5\le N_k\le 54.9$\\
\cline{2-4}
 & $0.25 \le w_{re}\le 1$ & $0.959\le n_s\le 0.966$ & $54.9\le N_k\le 67.4$\\
\hline
\multirow{2}{*}{$15 $} & $0 \le w_{re}\le 0.25$ & $0.956\le n_s\le 0.963$ & $46.6\le N_k\le 54.9$\\
\cline{2-4}
 & $0.25 \le w_{re}\le 1$ & $0.963\le n_s\le 0.969$ & $54.9\le N_k\le 67.4$\\
\hline
\multirow{2}{*}{$20 $} & $0 \le w_{re}\le 0.25$ & $0.957\le n_s\le 0.963$ & $46.6\le N_k\le 54.9$\\
\cline{2-4}
 & $0.25 \le w_{re}\le 1$ & $0.963\le n_s\le 0.970$ & $54.9\le N_k\le 67.5$\\
\hline
\end{tabular}
\caption{The allowed values of spectral index $n_s$ and number of efolds $N_k$ for various values of $X_0$ for inverse $\cosh$ 
potential considering   $T_{re}\ge100$GeV.} 
\label{table:coshnsnk}
\end{table}

It can be seen from Table.~\ref{table:coshnsnk} that for physically plausible values of $w_{re}$ i.e between $0$ and $0.25$ the number of efolds should have values between $N_k=46$ to $N_k=55$. If we consider $w_{re}\le 1$, we can allow $N_k$ to be around $67$. 
These upper bounds on $N_k$ and $n_s$ can be transferred into lower bounds on tensor-to-scalar ratio $r$.

The tensor-to-scalar ratio (\ref{tsratio}) for this model can be obtained from \eqrftw{ep1cosh}{xknk} in terms of $N_k$ as

\be
r=\frac{16}{X_0^2}\left(\frac{1}{\sinh\left(\frac{2 N_k}{X_0^2}\right)}\right),\label{coshtsratio}
\ee  
and the spectral index can also be expressed in terms of $N_k$ by inverting \eqrf{nkns} as
\be
n_s=1-\frac{2}{X_0^2}\frac{1}{\tanh\frac{N_k}{X_0^2}}.\label{nsnk}
\ee

\begin{figure}[h!]
 \centering
\subfigure{\includegraphics[width=7cm, height = 7cm]{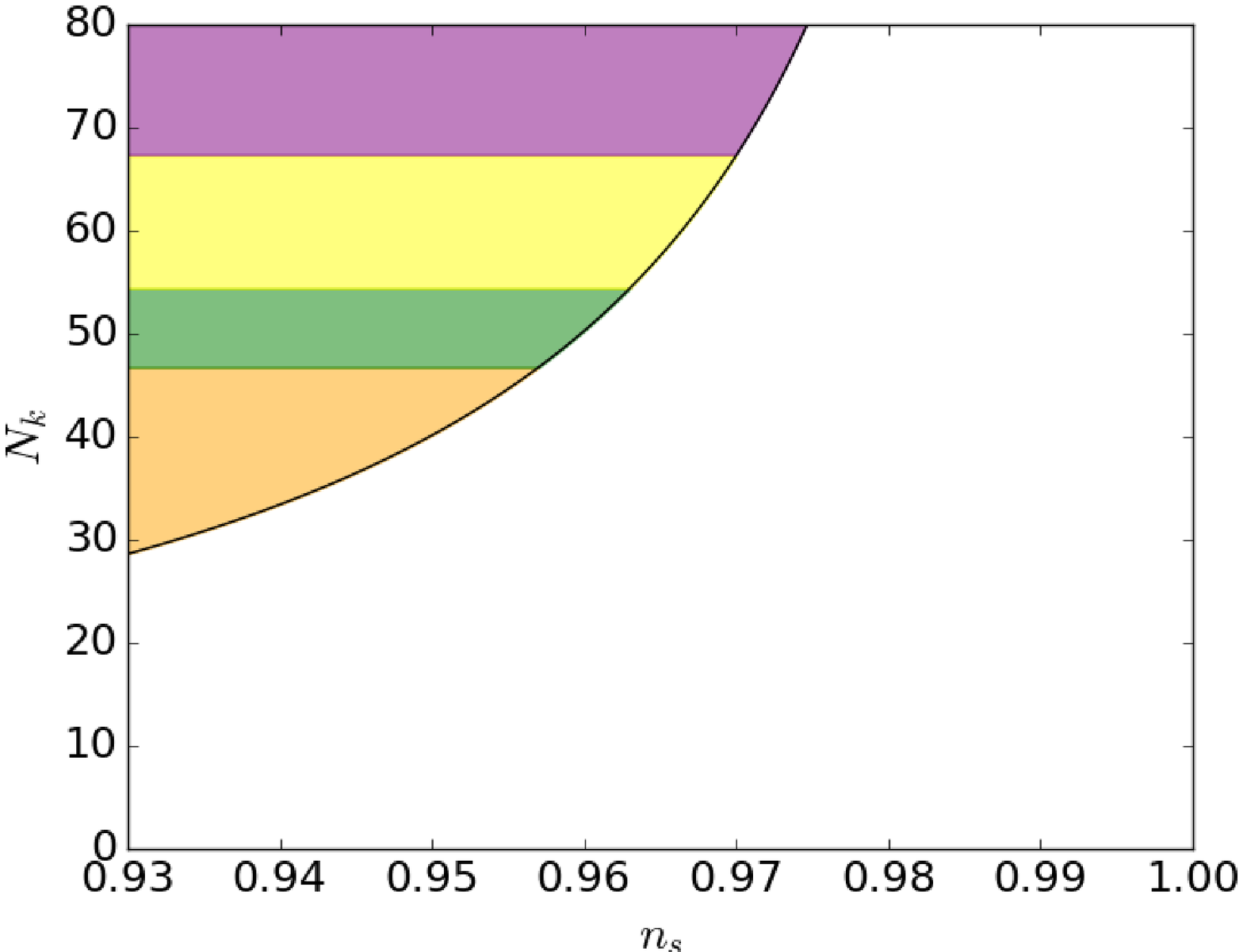}}
\subfigure{\includegraphics[width=7cm, height = 7cm]{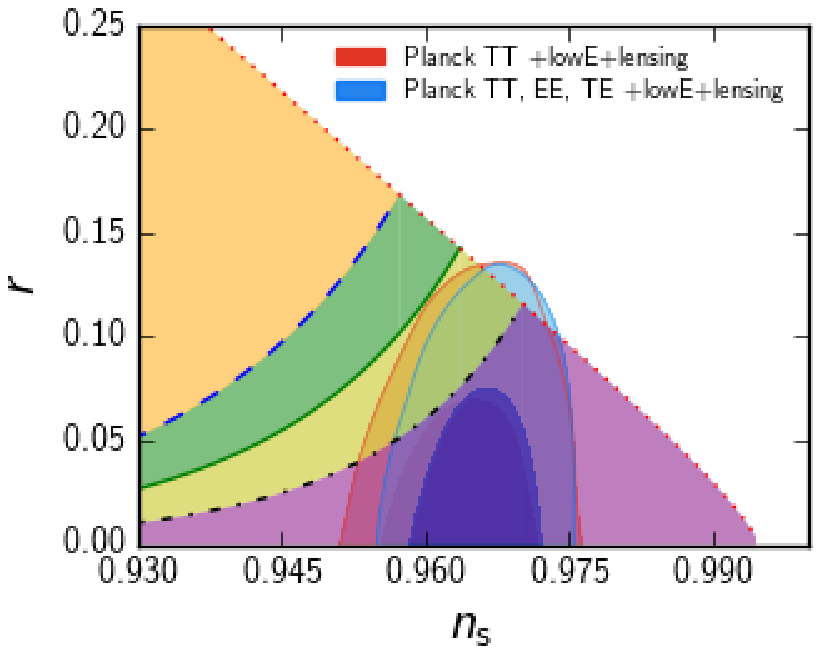}}
 \caption{ $N_k$ vs $n_s$ and $r$ vs $n_s$ 
predictions for inverse $\cosh$ potential along with  joint $68\%$CL and $95\%$CL Planck-2018 constraints. In both figures  
the orange region corresponds to $w_{re} < 0$, the green region corresponds to 
$0 < w_{re}< 0.25 $, the yellow region corresponds to $0.25 < w_{re} < 1$ and the purple region corresponds to $w_{re}>1$.
In the right panel of the figure 
dashed blue lines corresponds to $N_k=46$, solid green lines corresponds to $N_k=55$ and dashdotted black 
lines corresponds to $N_k=67$. These values of $N_k$ corresponds to bounds on $n_s$ obtained by demanding 
$T_{re}>100$GeV for different values of $w_{re}$ (see Table~\ref{table:coshnsnk}).
The solid black line in the left panel of the figure and dotted red line in the right panel of the  figure 
corresponds to $X_0=20$ and the colored
region corresponds to $X_0<20$. The colored region  extends further only slightly by increasing $X_0$.}
 \label{fig:coshrnsplanck}
\end{figure}

The predictions for $r$ and $n_s$ can be obtained for various values of $X_0$ and $N_k$ using \eqrftw{coshtsratio}{nkns} for 
the potential \eqrf{coshpt}. We chose the values of $N_k$ as $46$, $55$ and $67$ obtained by using bounds on $w_{re}$ 
(see Table.~\ref{table:coshnsnk}). 
Fig.~\ref{fig:coshrnsplanck} shows $N_k$ and $r$ as a function of $n_s$ corresponding to 
different values of equation of state during reheating $w_{re}$  along with joint $68\%$CL and $95\%$CL Planck-2018 constraints. 
It can be seen from Fig.~\ref{fig:coshrnsplanck} that the physically plausible value of the equation of 
state $0\le w_{re} \le 0.25$, which corresponds to $46\le N_k \le 55$ is disfavored by Planck observations and $w_{re}$ for 
these models should be close to $1$ to satisfy Planck-2018 constraints 
on $r$ and $n_s$. For example, if we put $N_k = 67$ (corresponding to $w_{re}=1$) and $X_0 = 8.3$ in 
\eqrftw{coshtsratio}{nsnk}, we obtain $r=0.068$ for $n_s=0.9612$ which are slightly within the 
Planck-2018 $68\%$CL value of $n_s=0.9653\pm 0.0041$, Planck-2018 $95\%$CL bounds on $r\le 0.07$ and slightly above 
the joint BICEP2/Keck Array and Planck bounds on $r\le0.06$ \cite{Ade:2018gkx}. If we decrease the value of $X_0$ further and/or
decrease the value $N_k$, the values of $n_s$ and $r$ predicted by \eqrftw{coshtsratio}{nsnk} move outside the Planck-2018 bounds.
The value of $X_0$ increases as we move from left bottom to the right top in the right panel 
of the Fig.~\ref{fig:coshrnsplanck} and the red dotted line in this panel corresponds to $X_0=20$.
From this analysis we find that for $n_s$-$r$ predictions of tachyon inflation with inverse $\cosh$ potential to fall well 
within Planck-2018 $1\sigma$ bounds and  BICEP2/Keck Array bounds on $n_s$ and $r$  one requires the equation of state 
$w_{re}$ to be larger than $1$, which violets causality.  

\subsection{Exponential potential}  Another string theory motivated potential for tachyon inflation is the 
exponential potential (\ref{exppotential}), which was studied by \cite{Sen:2002an,Sami:2002fs}. In terms of variable 
$x\equiv\frac{T}{T_0}$ it can be expressed as
\be
V(x)=\lambda e^{-x} \label{exppt}
\ee
The slow-roll parameters for this model can be expressed using \eqrftw{ep1}{ep2} as
\be
\epsilon_1=\frac{\epsilon_2}{2}=\frac{1}{2X_0^2}e^x. \label{expepsilon}
\ee
To find the value of tachyon field at the end of inflation we put $\epsilon_1=1$ and we get
\be
x_{end}=\ln\left(2X_0^2\right)\label{expxend}
\ee
Using \eqrf{efoldsT} the number of e-foldings $N_k$ for this potential can be obtained as
\be
N_k=X_0^2=\left(e^{-x_k}-e^{-x_{end}}\right)=X_0^2=\left(e^{-x_k}-\frac{1}{2X_0^2}\right) \label{expnk}
\ee 
One can see from this equation that $X_0^2\ge (N_k+\frac12)\ge N_k$. This is in contrast to the inverse $\cosh$ potential, where
sufficient number of e-foldings can be obtained with any value of $X_0$. The value of the tachyon field at the time when 
the mode $k$ leaves the inflationary horizon can be obtained using \eqrf{expnk} as
\be
x_k=\ln\frac{X_0^2}{N_k+\frac12}.\label{expxk}
\ee 
The spectral index $n_s$  for this model is expressed using \eqrftw{ns}{expepsilon} as
\be
n_s=1-\frac{2}{X_0^2}e^{x_k}, \label{expns}
\ee
The relation between $n_s$ and $N_k$ can be obtained using \eqrftw{expxk}{expns}  as
\be
N_k=\frac{2}{1-n_s}-\frac{1}{2}.\label{expnkns}
\ee 
The value of the potential at the end of inflation for this case can be expressed as
\be
V_{end}=\lambda e^{-x_{end}}=3M_p^2H_k^2\frac{e^{-x_{end}}}{e^{-x_k}},
\ee
which using \eqrftw{expxend}{expns} becomes
\be
V_{end}=\frac34 M_p^2 H_k^2\left(1-n_s\right).\label{expvendfinal}
\ee
Hubble constant at time of horizon exit of mode $k$ can be expressed in terms of scalar amplitude $A_s$ and spectral index $n_s$
using \eqrftr{scalaramp}{expepsilon}{expns} as 
\be
H_k=\pi M_p\sqrt{4A_s\left(1-n_s\right)\left(1-\frac{1}{12}\left(1-n_s\right)\right)}.\label{exphk}
\ee

\begin{figure}[h!]
 \centering
\includegraphics[width=7cm, height = 7cm]{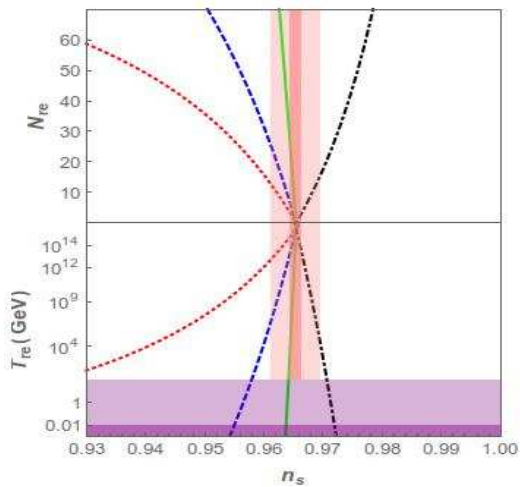}
\caption{Figure shows $N_{re}$ and $T_{re}$, the length of reheating and temperature at the end of reheating respectively, 
as a function of $n_s$  for exponential potential. Here all curves and shaded regions are same as Fig:~\ref{fig:coshnretre}.}
\label{fig:exptrenre}
\end{figure}

Again for the potential (\ref{exppt}), $N_k$, $V_{end}$ and $H_k$ are expressed in terms of $A_s$ and $n_s$ and one can obtain 
the temperature at the end of reheating $T_{re}$ and the number of efolds during reheating $N_{re}$ as a function of $n_s$ using 
\eqrftw{nrefinal}{trefinal}, which are shown in Fig.~\ref{fig:exptrenre}. As in the case of inverse $\cosh$ potential we have again
chosen four values of $w_{re}$ between $-\frac13$ to $1$. For the physically plausible value of $w_{re}$ i.e. 
$0\le w_{re} \le 0.25$ and $T_{re}\ge 100$GeV, the value of $n_s$ is restricted between $0.958\le n_s \le 0.964$.  This
again corresponds to $47\le N_k \le 55$. If we chose $0 \le w_{re} \le 1$, it can be seen from the Fig.~\ref{fig:exptrenre} that
$0.958\le n_s \le 0.971$ for $T_{re}\ge 100$GeV, which gives $47\le N_k \le 68$.

\begin{figure}[h!]
 \centering
\subfigure{\includegraphics[width=7cm, height = 7cm]{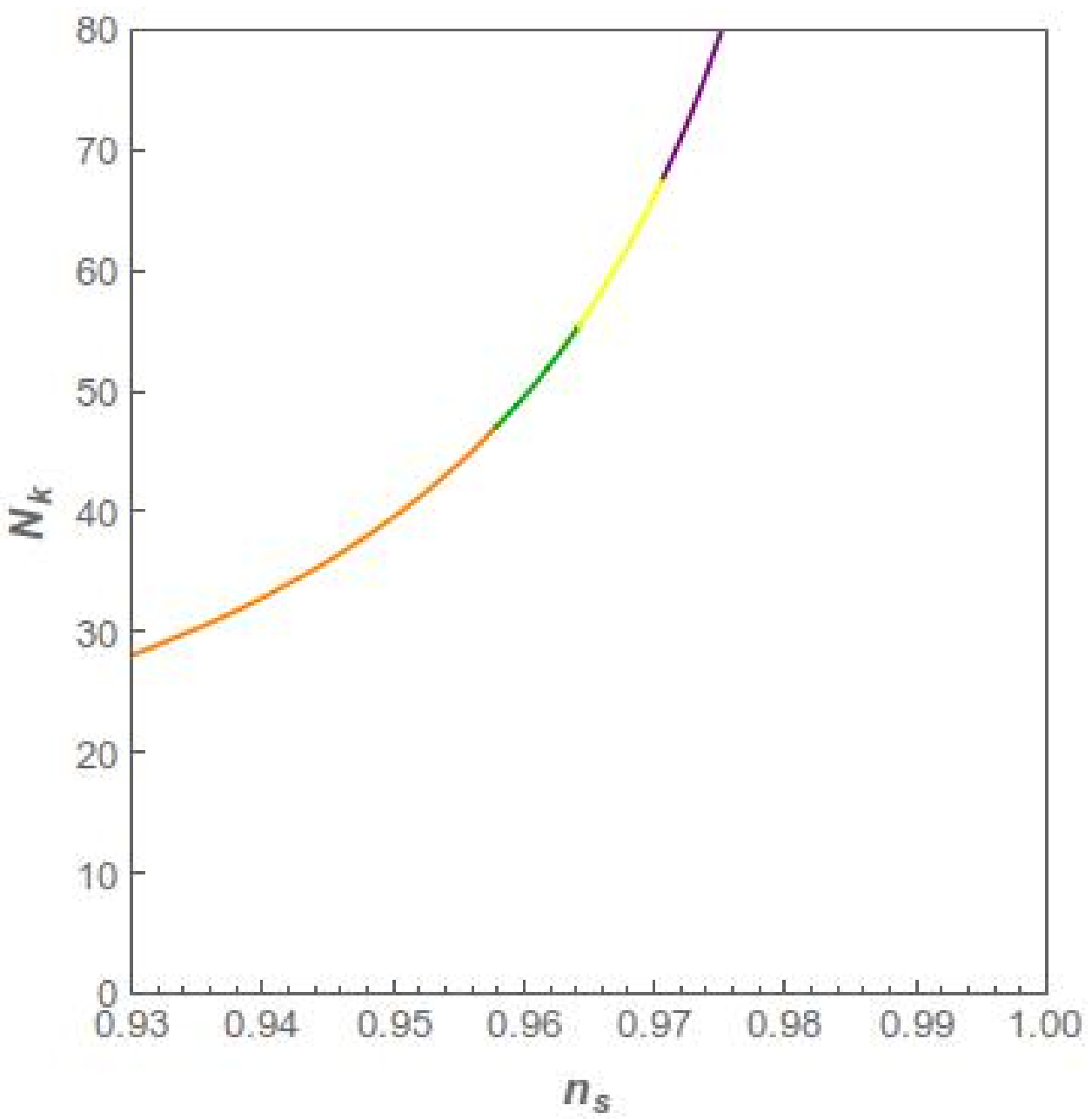}}
\subfigure{\includegraphics[width=7cm, height = 7cm]{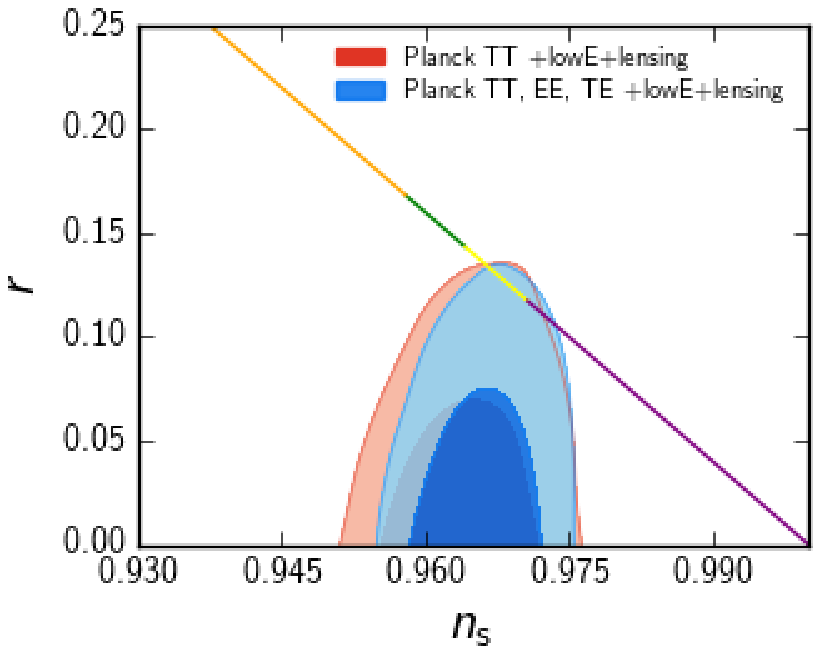}}
 \caption{$N_k$ vs $n_s$ and  $r$ vs $n_s$ predictions for exponential inflation along with joint $68\%$CL and $95\%$CL 
Planck-2015 constraints. 
In both the figures orange portion of the curve represents
allowed $r$ and $n_s$ for $w_{re}<0$, green portion corresponds to $w_{re}$ between $0$ and $0.25$, yellow portion corresponds to 
$w_{re}$ between $0.25$ and $1$ and purple portion corresponds to $w_{re}>1$. These bounds on $w_{re}$ corresponds to bounds on
$n_s$ and hence bounds on $N_k$ obtained by demanding $T_{re}>100$GeV. It can be seen from the figure that this model is
ruled out by Planck observations at $2\sigma$ for physically allowed equation of state during reheating $0 \le w_{re} \le 0.25$. }
 \label{fig:exprns}
\end{figure}

The tensor-to-scalar ratio (\ref{tsratio}) for this potential is given as
\be
r=\frac{8e^{x_k}}{X_0^2}=4\left(1-n_s\right),\label{exptsratio}
\ee
where we have used \eqrf{expns} in the last step. The plots for $N_k$ and $r$ as a function of $n_s$ are shown along with 
joint $68\%$CL and $95\%$CL Planck-2015 constraints in Fig.~\ref{fig:exprns}. The bounds on $n_s$ obtained by imposing the 
condition 
$0\le w_{re} \le 0.25$ and $T_{re}\ge 100$GeV provide bounds on $r$ as $0.144\le r \le 0.168$. If we consider the broader range
for $w_{re}$ i.e. $0\le w_{re} \le 1$, the bounds on $r$ become 
$0.116 \le r \le 0.168$, which is  above  than the Planck-2018  bound $r\le 0.07$ \cite{Akrami:2018odb} and the joint
BICEP2/Keck Array and Planck bounds on $r\le 0.06$ \cite{Ade:2018gkx}. As depicted in 
Fig.~\ref{fig:exprns}, the effective equation of state during reheating $w_{re}$ for this choice of potential 
should lie between $0.25$ and $1$ to satisfy Planck-2018 $95\%$CL constraints on $r$-$n_s$.
Hence  tachyon 
inflation with exponential potential (\ref{exppt}) is disfavored if physically plausible value of reheating equation of state
$0\le w_{re} \le 0.25$ is considered.

\section{Conclusions}\label{conclusions}
Tachyon inflation \cite{Gibbons:2002md,Sen:1999md,Garousi:2000tr,Bergshoeff:2000dq,Kluson:2000iy,Sen:2002qa,Kutasov:2003er,
Okuyama:2003wm} is  one of the most attractive models of $K$-inflation \cite{ArmendarizPicon:1999rj,Garriga:1999vw} 
motivated by string theory. 
In this work we have analyzed tachyon inflation by imposing constraints from reheating. This technique was earlier used to 
constrain various models of canonical inflation \cite{Dai:2014jja,Munoz:2014eqa,Cook:2015vqa}. 
Here we chose inverse $\cosh$ potential \cite{Sen:1999md, Garousi:2000tr, Lambert:2003Ar} and exponential potential 
\cite{Sen:2002an,Sami:2002fs} for our analysis.  
We compute reheating temperature $T_{re}$ and number of e-folds during reheating $N_{re}$ 
 as a function of spectral index $n_s$ for these potentials by assuming the effective equation of state during reheating $w_{re}$ 
to be constant. $w_{re}$ was obtained for various reheating scenarios \cite{Podolsky:2005bw} and it was found that 
$0\le w_{re} \le 0.25$. 
By demanding $0\le w_{re} \le 0.25$ and $T_{re} \ge 100$GeV we find bounds on $n_s$ and number of e-folds $N_k$
from the time when mode $k$ corresponding to pivot scale $k_0=0.05$Mpc\textsuperscript{-1} leaves inflationary horizon to the end
of inflation. These bounds restrict the allowed regions in $n_s$-$r$ plane for these potential. 

For inverse $\cosh$ potential (\ref{coshpotential}),
as shown in Fig.~\ref{fig:coshrnsplanck}, we find that $N_k$ should lie between $46$ and $55$ for $0\le w_{re} \le 0.25$. If we 
choose a broader range $0\le w_{re} \le 1$, $N_k$ can lie between $46$ and $67$. The $n_s$-$r$ predictions for inverse $\cosh$ 
potential lie outside the Planck-2018 bounds for physically plausible values $0\le w_{re} \le 0.25$.  

For exponential potential (\ref{exppotential}), the condition $0\le w_{re} \le 0.25$ and $T_{re} \ge 100$GeV gives bounds on 
$n_s$ as $0.958\le n_s \le 0.964$, which corresponds to $47\le N_k \le 55$. For $0\le w_{re} \le 1$ we obtain 
$47 \le N_k \le 68$. As shown in Fig.~\ref{fig:exprns}, for this model also, 
the $n_s$-$r$ predictions lie outside the Planck-2018 bounds for $0\le w_{re} \le 0.25$. We also find that $r\ge 0.116$ for 
this model for $w_{re} \le 1$, which is  higher than the Planck-2018 bound $r\le 0.1$\cite{Akrami:2018odb} and the 
joint BICEP2/Keck Array and Planck bound $r\le 0.06$ \cite{Ade:2018gkx}. Both exponential
potential and inverse $\cosh$ potential are disfavored by Planck-2018 bounds on $n_s$-$r$ for the physically plausible values 
of effective equation of state during reheating $0\le w_{re} \le 0.25$. For both these models $w_{re}$ close to $1$ is required 
to satisfy Planck bounds on $n_s$ and $r$. With tachyon potentials derived from string theory reheating is not well understood
\cite{Cline:2002it}. So this work can be helpful in determining correct mechanism for reheating in tachyon inflation.

\end{document}